# Multi-beam Energy Moments of Multibeam Particle Velocity Distributions

M. V. Goldman,[1] D. L. Newman,[1] J. P. Eastwood[2] and G. Lapenta[3]

[1]Physics Department, University of Colorado at Boulder.

[2]The Blackett Laboratory, Imperial College London, London SW7 2AZ, UK.

[3]Department of Mathematics, KU Leuven, University of Leuven, Celestijnenlaan 200B, 2001, Leuven, Belgium.

Martin V. Goldman ([goldman@colorado.edu](goldman@colorado.edu))

**Key Points:**

- *Standard* velocity moments of effectively-disjoint measured multibeam particle velocity distributions, f($\mathbf{v}$) can be misleading.

- So-called *multibeam moments* of multibeam f($\mathbf{v}$) are found by taking a standard moment of each beam and then summing over beams.

- Standard thermal moments (pressure, thermal flux, etc.) of multibeam f($\mathbf{v}$) can contain "*false*" parts which are absent in multibeam moments.

**Abstract**

High resolution electron and ion velocity distributions, f($\mathbf{v}$), which consist of N effectively disjoint beams, have been measured by NASA's Magnetospheric Multi-Scale Mission (MMS) observatories and in reconnection simulations. Commonly used *standard* velocity moments generally assume a single *mean-flow-velocity* for the entire distribution, which can lead to counterintuitive results for a multibeam f($\mathbf{v}$). An example is the (false) standard thermal energy moment of a pair of equal and opposite cold particle beams, which is nonzero even though each beam has zero thermal energy. By contrast, a *multibeam* moment of two or more beams has no false thermal energy. A multibeam moment is obtained by taking a standard moment of each beam and then summing over beams. In this paper we will generalize these notions, explore their consequences and apply them to an f($\mathbf{v}$) which is sum of tri-Maxwellians.

Both standard and multibeam energy moments have coherent and incoherent forms. Examples of *incoherent* moments are the thermal energy density, the pressure and the thermal energy flux (enthalpy flux plus heat flux). Corresponding *coherent* moments are the bulk kinetic energy density, the RAM pressure and the bulk kinetic energy flux. The false part of an incoherent moment is defined as the difference between the standard incoherent moment and the corresponding multibeam moment. The sum of a pair of corresponding coherent and incoherent moments will be called the *undecomposed* moment. *Undecomposed* moments are independent of whether the sum is *standard* or *multibeam* and therefore have advantages when studying moments of measured f($\mathbf{v}$).

# 1 Introduction

## 1.1 VELOCITY MOMENTS OF MULTIBEAM VELOCITY DISTRIBUTIONS, f($\mathbf{v}$)

In space plasma physics is it very common to discuss the properties of a plasma in terms of its bulk properties – density, velocity, pressure, etc. and to use a fluid theory framework to describe and predict the behavior of many space plasma phenomena (Cravens, T.E.;  Bellan, P.;  Kulsrud, R.M..).  Two different kinds of space plasmas are commonly treated as fluids: collisional and collisionless.

- In collisional space plasmas, such as the lower solar corona and lower ionosphere, the collisional mean free path is short and the particle velocity distributions, f($\mathbf{v}$), are at or close to Maxwellian (e.g., thermal equilibrium).  In this case the fluid equations for a given species can be derived from the Boltzmann equation for that species by the Chapman-Enskog procedure (Braginskii) and include collisional transport coefficients such as electron-ion equilibration rates and thermal conductivity.

- However, in Earth's upper ionosphere, magnetosphere and solar wind, the collisional mean free path is large, so the plasma is *collisionless* and often *far* from thermal equilibrium [Pashmann et al 1998]. The velocity distributions, can have nonthermal tails and may even be multipeaked, leading to effectivey disjoint f($\mathbf{v}$) [Burch, et al].  *Standard* collisionless fluid equations and energy transport equations can be derived for constituent species by taking velocity moments of the Vlasov eqn. for each species (Bellan, 2006; Aunai, et al, ;  Goldman, et al, 2015).  The moment equations are expressed in terms of (standard) velocity moments of the particle distribution, f($\mathbf{v}$).  In collisionless space-plasma fluid equations there are no electron-ion collisions and there is no collisional conductivity. Energy transport can only occur by *convection*, described in terms of particle energy fluxes or by radiation (Poynting flux).

Other terms in the *standard* collisionless fluid equations resemble those in the collisional fluid equations even though f($\mathbf{v}$) is not generally Maxwellian.  For a given f($\mathbf{v}$) there is one flow coherent velocity, $\mathbf{u}$ and one mean density, n; higher-order moments decompose into a sum of a cohererent part (involving $\mathbf{u}$) and an incoherent ("thermal") corresponding to moments of velocity fluctuations, such as $\int d^3 \mathbf{v} f(\mathbf{v})|\mathbf{v} - \mathbf{u}|^2$ or  $\int d^3 \mathbf{v} f(\mathbf{v}) \, (\mathbf{v} - \mathbf{u}) \, |\mathbf{v} - \mathbf{u}|^2$.  Thus, the energy density of a given species is the sum of a bulk kinetic energy density, $nmu^2/2$, and a "thermal" energy density moment, $m\int d^3 \mathbf{v} f(\mathbf{v})|\mathbf{v} - \mathbf{u}|^2/2$; the energy density *flux* is a sum of a coherent bulk kinetic energy flux, $\mathbf{u}nmu^2/2$, and a "thermal" energy flux moment (enthalpy plus heat flux).

The meaning of each of the decomposed moments is well understood and not controversial for contiguous and effectively single-peaked velocity distributions, f($\mathbf{v}$).  However, for multibeam (effectively disjoint) f($\mathbf{v}$), standard velocity moments can give rise to misinterpretations such as false identifications of incoherent energy density and temperature.  In this paper an alternate method is developed for taking velocity moments of an N-beam velocity distribution, f($\mathbf{v}$), of form f($\mathbf{v}$) = $f_1(\mathbf{v}) + f_2(\mathbf{v}) + \ldots + f_N(\mathbf{v})$.  Such *multibeam* moments are formed by taking standard coherent moments or standard thermal moments of each beam, $f_j(\mathbf{v})$ and then summing over beams j = 1 to N.  This procedure eliminates *false thermal parts* of higher-order standard decomposed moments of f($\mathbf{v}$), such as the standard "thermal" energy density and related false temperatures.  However, it is important to note that the sum of the coherent plus the "thermal" parts of a higher-order moment remains the same whether the velocity moments are standard or multibeam.  Such a sum will be called an *undecomposed* higher-order moment.  An example is the undecomposed particle energy density, $nm<|\mathbf{v}|^2>/2$, which can be decomposed by writing $\mathbf{v} = \mathbf{u} + (\mathbf{v} - \mathbf{u})$.

In practical space plasma applications $f(\mathbf{v})$ is measured by electrostatic analyser instruments [Fazakerley et al., 1998], and (standard) moments can be constructed on the ground or onboard (so reducing the telemetry) (Reme et al. Ann Geophys 2001;  McFadden et al. 2007) The launch of the Fast Plasma Instrument (FPI) (Pollock et al., 2016) on board the Magneto-sphere MultiScale (MMS) Mission [Burch, 2016] now enables the full 3D electron and ion distribution to be measured over unprecedented time-intervals of 30 ms and 150 ms cadence respectively (or 7.5 ms and 37.5 ms in certain operating modes [Rager, et al 2018]).

In many regions of interest, not least the electron dissipation region of magnetic recon-nection (e.g., Burch, et al, 2015), the ion and electron velocity distributions, $f(\mathbf{v})$, consist of a number of effectively disjoint pieces in velocity space, which we will refer to as *beams*.  This calls into question the use of standard moments because experimentally the distribution is not contiguous; nor is it single peaked. In particular, difficulties of interpretation may arise, and the underlying physics may become obscured by using a description which does not adequately capture the phenomena being observed.

Consequently, for multibeam $f(\mathbf{v})$ it may be more useful *not* to take moments in the *standard* way but rather to take standard decomposed moments beam by beam and then add them together.  We call such sums *multibeam* moments. The purpose of this paper is to lay out a systematic framework for understanding multibeam moments. By way of introduction, we first examine a highly idealized theoretical example which also serves to illustrate difficulties of interpretation that may arise.

## 1.2   FALSE THERMAL ENERGY MOMENT OF A PAIR OF COLD BEAMS

Fig. 1 shows two equal and opposite cold electron beams, with velocities $\mathbf{u}_0$ and $-\mathbf{u}_0$ and equal densities $n_0$. According to standard moment theory the effective velocity distribution, $f(\mathbf{v})$ is one entity, with one flow velocity, $\mathbf{u}$. In this example, $\mathbf{u} = 0$, so the *bulk kinetic energy moment,* $U_{bulk}$ is zero and the (single) density is $n = 2n_0$. The *incoherent* energy density moment found from *standard* moment theory is $mn(u_0)^2/2$. This incoherent part of the energy density is often called the *thermal energy density* and written as $U_{therm} = n"T"$. This yields an effective *temperature,* "T" = $m(u_0)^2/2$. A difficulty with such a *standard* moment is that the pair of *cold moving* beams appears to have a temperature, which we will call here a *false-temperature*, but *no* bulk kinetic energy density.

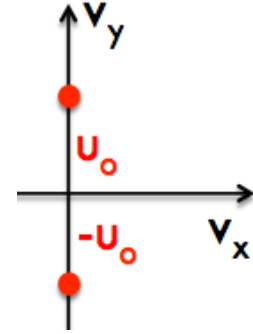

**Figure 1** Two equal and opposite cold beams

This is customarily remedied by simply considering $f(\mathbf{v})$ to be a *two-beam* system, $f(\mathbf{v}) = f_1(\mathbf{v}) + f_2(\mathbf{v})$. The density and velocity moments of $f_1$ and $f_2$ are $\{n_1, \mathbf{u}_1\}$ and $\{n_2, \mathbf{u}_2\}$, where, $n_1 = n_2 = n_0$, $\mathbf{u}_1 = \mathbf{u}_0$ and $\mathbf{u}_2 = -\mathbf{u}_0$. The energy moments of *each* beam are now $U_{bulk1,2} = n_0 m(u_0)^2/2$ and $U_{therm1,2} = 0$. For the *system* of two beams, $(U_{bulk})^{2\text{-beam}} = U_{bulk1} + U_{bulk2} = nm(u_0)^2/2$, where $n = n_1 + n_2 = 2n_0$ and $(U_{therm})^{2\text{-beam}} = 0 + 0$, which is more intuitive than the results of the standard moment analysis, $U_{bulk} = 0$ and $U_{therm} = nm(u_0)^2/2$.

We summarize the results for the two different ways of taking energy moments of $f(\mathbf{v}) = f_1(\mathbf{v}) + f_2(\mathbf{v})$ in the table below:

**Table 1:** Energy density moments for two equal and opposite cold (e or i) beams

| Standard energy moments: 2 cold beams | | Multibeam energy moments: 2 cold beams | |
|---|---|---|---|
| $U_{bulk}$ | 0 | $(U_{bulk})^{2\text{-beam}}$ | $nm(u_0)^2/2$ |
| $U_{therm}$ | $nm(u_0)^2/2$ | $(U_{therm})^{2\text{-beam}}$ | 0 |
| $U = U_{bulk} + U_{therm}$ | $nm(u_0)^2/2$ | $U = (U_{bulk})^{2\text{-beam}} + (U_{therm})^{2\text{-beam}}$ | $nm(u_0)^2/2$ |

The bottom (shaded) row demonstrates an important general result: that the *sum* of the bulk and thermal energy terms has the same value, $U = nm(u_0)^2/2$, for both the *standard* and the *multibeam* methods of taking moments. In this paper, the *generalized* U will be referred to as the *undecomposed* energy moment and the bulk and thermal energies as the decomposed energy moments. It is clear from Table 1 that the values of the decomposed moments differ, depending on the method used for taking moments. Since U is independent of the approach used, it can be employed reliably to quantify the properties of the plasma without ambiguity. This offers a path forward in studying the properties of complex distribution functions such as those observed by MMS.

### 1.3. ORGANIZATION AND OUTLINE OF THIS PAPER

In the present paper the concepts discussed above in Fig. 1 and Table 1 for two cold beams will be generalized to N-beams and the full set of multibeam pressure, energy and energy flux moments. Standard and multibeam moments then will be evaluated and compared for an analytic particle velocity distribution, f($\mathbf{v}$), consisting of a sum of N tri-Maxwellian beams.

The content of this paper is indicated below in terms of questions to be addressed:

1. Examples of measured and simulated f(v) *which are effectively disjoint.*

   - What are examples of effectively disjoint velocity distributions, f($\mathbf{v}$) ≡ f($\mathbf{v}$, $\mathbf{r}_0$, $t_0$), found at certain locations and times during magnetic reconnection in the magnetopause and found in PIC simulations of reconnection in the magnetotail?

   - What is a possible physical origin of such beams at {$\mathbf{r}_0$, $t_0$} in terms of their Lagrangian trajectories at other points in space and time?

2. Standard energy transport theory

   - What are the various energy moments and transport equations associated with a given f($\mathbf{r}$,$\mathbf{v}$,t)?

   - How are transport equations and transport coefficients derived from kinetic theory (i.e., the Vlasov equation) in a collisionless plasma?

3. Multibeam moments

   - What is the difference between *undecomposed* energy density, pressure and energy flux moments and *decomposed* energy density moments (bulk, thermal), pressure moments (RAM, thermal) and energy flux moments (bulk, enthalpy, heat flux)?

   - What is the relationship between *standard* and *multibeam* decomposed moments?

   - How are *false*-thermal parts of standard decomposed moments determined?

   - What are the *standard* and *multibeam* moments of a set of N tri-Maxwellian beams?

## 2. Measurements and simulations exhibiting multibeam velocity distributions

### 2.1 MULTIPLE ION BEAMS NEAR X-LINE DURING DAYSIDE RECONNECTION

In a pioneering paper (Burch, et al, 2016), high resolution data from the Fast Plasma Instrument (FPI) onboard MMS satellites was used to study electron and ion velocity distributions in the electron diffusion region of Earth's magnetopause during magnetic reconnection. The left image in Figure 2 shows the magnetic field and standard particle and field energy fluxes as functions of time along the trajectory of MMS4. The reconnection x-line is crossed at time 1307, as indicated by the vertical dashed line. The right image (here constructed from the data) shows 3D contour maps of the ion distribution, f(**v**), measured over a time interval of 150ms a few seconds before and a few seconds after an x-line crossing. Details are given in the Figure caption. It is clear that both ion distributions, f(**v**), are far from equilibrium and disjoint. Each f(**v**) is therefore an ion multibeam distribution amenable to the multibeam moment treatment described in this paper. Such a treatment will generally give ion energy moments *different* from the ones on the left in Figure (2).

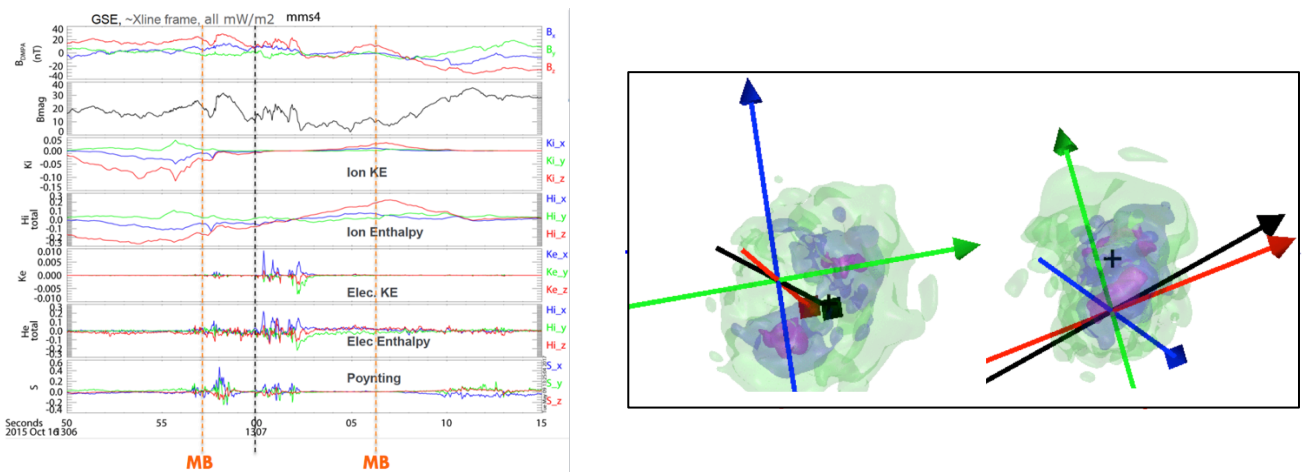

**Figure 2: Data and 3-D ion velocity distribution** measured by MMS-4 during the dayside EDR event studied in [Burch, et al, 2016]. Left panel: MMS4-measured particle energy and Poynting fluxes showing x-point around time 1307 (vertical black dashed line). The two vertical orange dotted lines labelled MB are at times ~3s before 1307 and ~6s after 1307. Multiple-beam-like features are found in the ion velocity distribution, f(**v**), at both of those times. The right two panels show f(**v**) determined by the Fast Plasma Instrument (FPI) at the later time from two different perspectives. In those panels the black arrow is in the direction of **B** and the blue, green and red arrows show the directions of the magnetopause GSE coordinates in x, y, z. The tip of each colored arrow is at a velocity corresponding to an energy of 1.6 keV. The values of f(**v**) on successive surface contours differ by factors of 5.

## 2.2 TAIL RECONNECTION <u>SIMULATION</u> SHOWS MULTIPLE ION BEAMS

Multibeam (disjoint) ion distributions, f($\mathbf{v}$), have also been found in *simulations* of magnetic reconnection in Earth's magnetotail. An example can be found in Eastwood, et al, in which PIC simulations are carried out in support of THEMIS observations of magnetic reconnection on February 27, 2009 where complex ion distributions were observed. The THEMIS magnetic field data were used to establish appropriate comparison cuts through a particle-in-cell simulation of reconnection, and very good agreement was found between the observed and simulated ion distributions on both sides of the dipolarization front.

The principal feature of interest in the simulation is the dynamics of a pair of counterpropagating ion beams found in the dipolarization front about five ion inertial lengths in front of the unperturbed plasma just below the neutral sheet at time t = 30. They are visible in the 2D reduced distribution in the $v_x$-$v_y$ (reconnection) plane shown in Figure 3a. This reduced distribution is integrated over $v_z$. The two beams are represented by two small cubes in velocity space labeled A and B. The other two orthogonal reduced distribution functions are shown in Figures 3b and 3c.

Figures 3d–3g show the real space self-consistent *trajectories* of the bunches of ions that pass through these two phase-space cubes at $t = 30$ over the time interval $23 < t < 35$. Trajectories are of particular interest because they illustrate *how* multibeam particle distributions, f($\mathbf{v}$), might arise.

The trajectories are superposed on plots of $E_{zSIM}$ (approximately the reconnection electric field) and $E_{ySIM}$ (approximately the Hall electric field), respectively. The fields and circled groups of ions in Figures 3d and 3e correspond to the start time for the trajectories ($t = 23.28$). The fields and circled bunches of ions in Figures 3f and 3g correspond to the time when the trajectories cross, thereby giving rise to the disjoint reduced distribution function show in Figures 3a-c. The ions composing the two beams are clearly distinguishable, with beam A moving primarily downward and to the right, while beam B moves primarily upward and to the right (as labeled in Figure 3d).

At the start time, the ion bunches are widely separated in space, with each bunch near a different separatrix. The ion bunch trajectories later in time are determined by the electric and magnetic forces. The ions are not frozen-in (this is the ion-diffusion region). For details concerning the forces and their influence on the ion trajectories refer to Eastwood, et al.

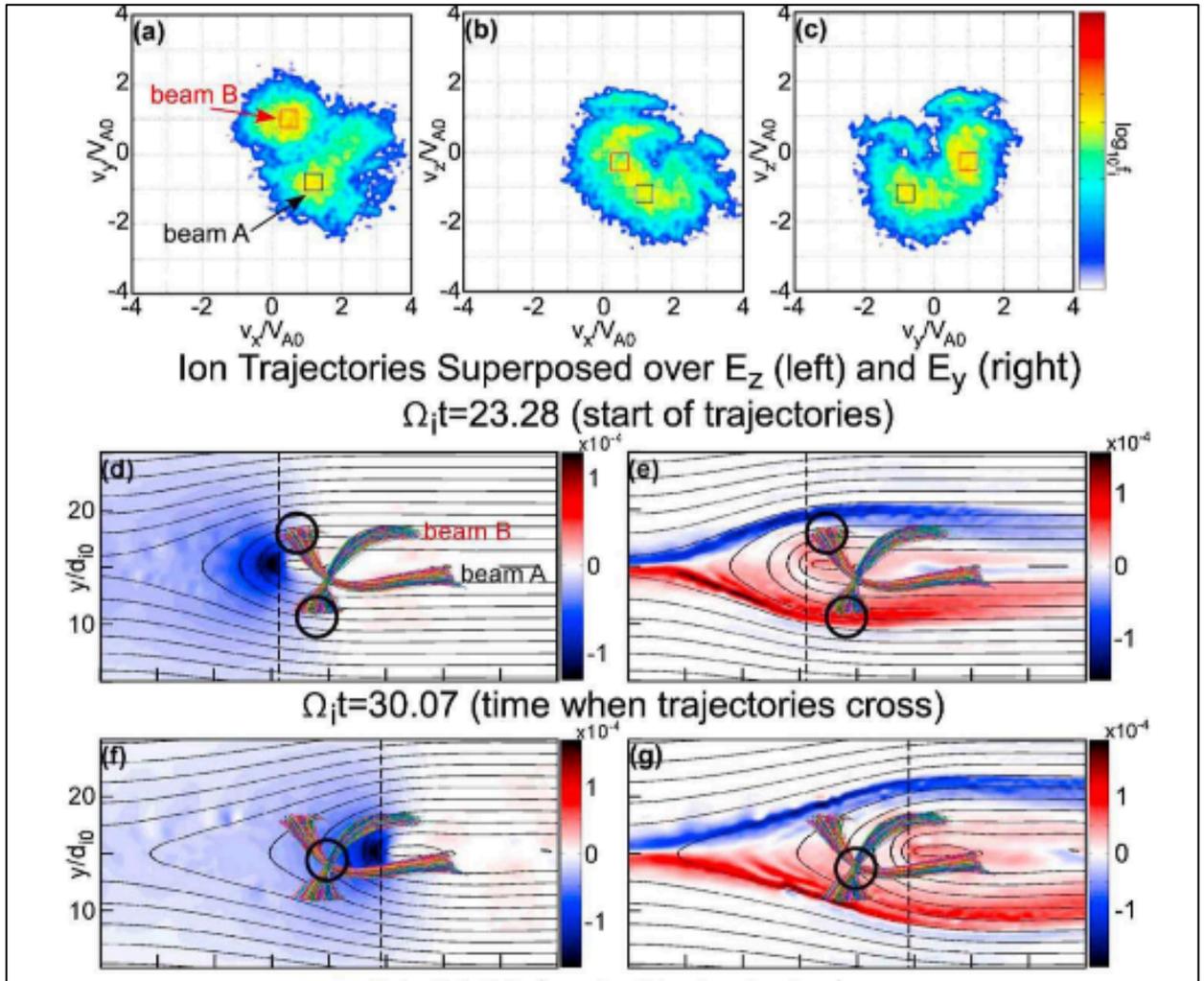

**Figure 3** Ion beams from PIC simulation from Eastwood, et al. (a–c) Ion beams observed in the ion distribution at $x$, $y$, and $t = 120$, $14$, and $30.07$ are shown. (d–h) Back and forward tracing analysis of the ion trajectories in space between $t = 23$ and $t = 35$, superimposed on snapshots of (d and f) $E_{z;SIM}$ and $E_{ySIM}$ at the times shown in the panel captions. For each snapshot, the physical locations of the ions in each beam along the overall trajectory are approximately indicated by the black circles.

# 3 "Standard" energy-transport theory

## 3.1 REVIEW OF STANDARD ENERGY-TRANSPORT THEORY

There are a number of different ways to take energy moments and derive corresponding energy transport equations for a given particle species. A common strategy, the so-called standard approach, is based on the assumption that there is a single mean-flow-velocity, **u** of a distribution f(**v**). For a given particle species the well-known equations (Bellan) for the transport of the bulk coherent energy density, $U_{bulk}$, and the "thermal" (incoherent) energy density, $U_{thermal}$ are given by

(1)
$$a) \quad \partial_t U_{bulk} + \nabla \cdot \mathbf{Q}_{bulk} = \mathbf{J} \cdot \mathbf{E} - \mathbf{u} \cdot \nabla \cdot \boldsymbol{P}$$
$$b) \quad \partial_t U_{thermal} + \nabla \cdot \mathbf{Q}_{thermal} = \mathbf{u} \cdot \nabla \cdot \boldsymbol{P}, \quad \mathbf{Q}_{thermal} \equiv \mathbf{Q}_{heatflux} + \mathbf{Q}_{enthalpy}$$

In Eqn. (1a), **u** is the flow-velocity of the species, $U_{bulk} = mnu^2/2$, $\mathbf{Q}_{bulk}$ is the bulk kinetic energy density flux, $\mathbf{u}U_{bulk}$, $\mathbf{J} \cdot \mathbf{E}$ is the work performed by the electric field, **E** against the current, $\mathbf{J}_e$ or $\mathbf{J}_i$. The term $-\mathbf{u} \cdot \nabla \cdot \mathbf{P}$ is the work performed by the pressure force, $-\nabla \cdot \mathbf{P}$, on the flow, **u**.

In Eqn. (1b), $\mathbf{Q}_{therm}$ is the thermal (incoherent) energy flux, $\mathbf{Q}_{therm} = \mathbf{Q}_{heatflux} + \mathbf{Q}_{enthalpy}$, consisting of a part, $\mathbf{Q}_{heatflux}$, which is invariant under velocity frame transformations and a part, $\mathbf{Q}_{enthalpy}$, proportional to **u**. The term $+\mathbf{u} \cdot \nabla \cdot \mathbf{P}$ is the work done by **u** against the pressure force. Hence, the work associated with pressure governs the transfer of energy between coherent and incoherent energy densities.

*Another* energy transport equation results from adding Eqn. (1a) to Eqn. (1b). The work terms associated with pressure forces then drop out entirely, resulting in:

(2a) $\quad \partial_t U + \nabla \cdot \mathbf{Q} = \mathbf{J} \cdot \mathbf{E}$

(2b) $\quad U \equiv U_{bulk} + U_{therm}, \quad \mathbf{Q} \equiv \mathbf{Q}_{bulk} + \mathbf{Q}_{therm}$

Eqn. (2) describes particle energy transport (for ions or electrons). We shall refer to Eqn. (2) as the *undecomposed* particle energy transport equation, and U as the undecomposed particle energy density. $U_{bulk}$ and $U_{thermal}$ can be understood as resulting from the decomposition of U into a coherent and an incoherent part. In a similar manner, **Q** is the undecomposed energy flux whose decomposed pieces, $\mathbf{Q}_{bulk}$ and $\mathbf{Q}_{thermal}$ are its coherent and incoherent parts.

At first glance energy transport Eqn. (2) appears less useful than energy transport Eqns. (1) because U and **Q** do not distinguish between coherent and incoherent (thermal) energies and fluxes. However, energy transport in terms of U and **Q** can be very useful because U and **Q** are independent of whether their decomposed parts are standard (single-beam) moments or multibeam moments of a given velocity distributions, f(**v**). Note also that U and **Q** enter into the electromagnetic energy transport equation in the same way as the electromagnetic energy density and Poynitng flux when $-(\mathbf{J}_e + \mathbf{J}_i) \cdot \mathbf{E}$ is eliminated by using Eqn. (2a) for each species:

(3)    $\partial_t \left( U_{EM} + U_e + U_i \right) + \nabla \cdot \left( \mathbf{S}_{Poynting} + \mathbf{Q}_e + \mathbf{Q}_i \right) = 0$

In Eqn. (3) the electromagnetic energy, $U_{EM} \propto E^2 + B^2$, and Poynting flux, $\mathbf{S} \propto \mathbf{E} \times \mathbf{B}$, are on the same footing as $U_{e,I}$ and $Q_{e,i}$. From Eqn. (3) it is evident that when magnetic energy is lost locally it can transform into *local* particle energy densities, $U_{e,i}$, and/or it can be transported *remotely* by radiation (Poynting flux) and convection (particle fluxes, $\mathbf{Q}_e$ and $\mathbf{Q}_i$).

At the level of description of Eqn (3) one cannot determine whether lost magnetic energy goes mainly into local particle heating, into particle acceleration or is spatially transported by radiation or coherent or incoherent particle convection. To determine in standard transport theory whether coherent or incoherent processes dominate changes in particle energy (or energy flux) one must employ the standard decomposed energy transport Eqns. (1).

### 3.2 KINETIC THEORY OF STANDARD ENERGY TRANSPORT

Further insight into the meaning of Eqn. (2) is gained by deriving it from kinetic theory (Aunai, et al; Bellan; Goldman, et al) In kinetic theory, undecomposed energy transport is described by taking a kinetic energy velocity moment of the Vlasov equation for either species:

$$\int d^3 \mathbf{v} \left[ \left( \frac{m_s v^2}{2} \right) \cdot \left\{ \partial_t + \mathbf{v} \cdot \nabla + \frac{q_s}{m_s} \left( \mathbf{E} + \frac{\mathbf{v}}{c} \times \mathbf{B} \right) \cdot \partial_\mathbf{v} \right\} f_s \right] = 0$$

The result is Eqn. (2) with the undecomposed U and $\mathbf{Q}$ now explicitly defined as velocity moments of f(**v**),

4a)   $U\left(\mathbf{r},\mathrm{t}\right) = $ *undecomposed* energy (e or i) $\equiv \int d^3 \mathbf{v} f\left(\mathbf{v}, \mathbf{r}, \mathrm{t}\right) \left[ \frac{mv^2}{2} \right]$

4b)   $\mathbf{Q}\left(\mathbf{r},\mathrm{t}\right) = $ *undecomposed* energy flux $= \int d^3 \mathbf{v} f\left(\mathbf{v}, \mathbf{r}, \mathrm{t}\right) \cdot \left[ \mathbf{v} \frac{mv^2}{2} \right]$

In the usual standard approach employing the mean-flow-velocity decomposition of U and $\mathbf{Q}$ moments, one substitutes $\mathbf{v} = \mathbf{u} + \delta\mathbf{v}$ into $mv^2/2$ and $m\mathbf{v}v^2/2$ in the velocity integrals in Eqns (4a) and (4b). Integrals whose integrands are linear in $\delta\mathbf{v} = (\mathbf{v}\text{-}\mathbf{u})$ vanish — hence the term "incoherent" for moments proportional to $(\delta\mathbf{v})$.[2] This kind of decomposition is most meaningful for a single beam (f(**v**) single-peaked and contiguous). The mean-flow-velocity decomposition of U and $\mathbf{Q}$ moments is

5a)   $U_{bulk} = \frac{nmu^2}{2}$,   $U_{thermal} = \frac{m}{2} \int d^3 \mathbf{v} f\left(\mathbf{v}\right)\left(\delta v\right)^2$

*5b)* $\quad \mathbf{Q}_{bulk} = n\mathbf{u}\dfrac{mu^2}{2}$

*5c)* $\quad \mathbf{Q}_{heatflux} = \displaystyle\int d^3\mathbf{v}f(\mathbf{v})\dfrac{m}{2}\left[\delta\mathbf{v}(\delta\mathbf{v})^2\right]$

*5d)* $\quad \mathbf{Q}_{enthalpy} = \mathbf{u}\cdot\displaystyle\int d^3\mathbf{v}f(\mathbf{v})\dfrac{m}{2}\left[(\delta\mathbf{v})^2 + 2\delta\mathbf{v}\delta\mathbf{v}\right]$

The three fluxes, **Q**, in Eqn. (5b-d) correspond to standard energy transport by convection of coherent and incoherent particle energies.

The enthalpy flux is second-order in $\delta v$ so it may be defined in terms of the pressure tensor, **P**. In dimensional (as opposed to dimensionless) units,

*5e)* $\quad \mathbf{Q}_{enthalpy} = \dfrac{\mathbf{u}\,\mathrm{Tr}(\mathbf{P})}{2} + \mathbf{u}\cdot\mathbf{P}, \quad \mathbf{P} \equiv m\displaystyle\int d^3\mathbf{v}f(\mathbf{v})\delta\mathbf{v}\delta\mathbf{v}$

The moments in Eqn. (5) enter into the two independent energy transport Eqns. (1a,b) for $U_{bulk}$ and $U_{therm}$. They can be derived from the Vlasov Eqn (Bellan; Goldman, et al) by using

- the transport Eqn (2a) for $U = U_{bulk} + U_{therm}$ (as found above from the Vlasov Eqn.) and
- a transport equation for $U_{bulk}$ obtained by dotting u into the *momentum* eqn (*first* moment of the Vlasov eqn.).

The distinction between undecomposed and decomposed moments may be applied to the pressure tensor moment as well as to the energy and energy flux moments. The *stress tensor*, **T**, is the *undecomposed* moment which decomposes into a *RAM pressure* tensor (dyadic) and a *"thermal" pressure*, **P** when the mean velocity, **u** is introduced:

*6)* $\quad \mathbf{T} \equiv m\displaystyle\int d^3\mathbf{v}f(\mathbf{v})\mathbf{v}\mathbf{v} = \mathbf{P}_{RAM} + \mathbf{P}, \quad \mathbf{P}_{RAM} = nm\mathbf{u}\mathbf{u}$

Note that the pressure tensor, **P**, is frame-invariant (like $U_{therm}$ and $\mathbf{Q}_{heatflux}$), whereas **T** and $\mathbf{P}_{RAM}$ are not. In *dimensional units,*

*6a)* $\mathrm{Tr}(\mathbf{P})/2 = U_{therm}$

This helps explain the enthalpy flux in Eqn. (5e), since the first term on the right is now $\mathbf{u}U_{therm}$. The undecomposed stress tensor enters directly into the force-momentum equation, which is the *first* moment of the Vlasov equation and may be written in the following form:

*6b)* $\quad \partial_t(mn\mathbf{u}) + \nabla\cdot\boldsymbol{T} = q\mathbf{E} + q\dfrac{\mathbf{u}}{c}\times\mathbf{B}$

When **T** is decomposed the "thermal" pressure force appears explicitly and the RAM pressure leads to the convective derivative, $\mathbf{u}\cdot\nabla\mathbf{u}$.

A space-time transport eqn for **T**(r,t) is generated from kinetic theory by multiplying the Vlasov eqn by m**vv** and integrating over **v**. However, space-time transport will not be studied in this paper.

### 3.3 STANDARD MOMENTS OF ONE TRI-MAXWELLIAN BEAM, f(**v**)

To give explicit examples of standard and multibeam energy moments, we first consider a single tri-Maxwellian distribution function, f(**v**), of form,

$$(7) \quad f(\mathbf{v}) = \frac{n}{w_x w_y w_z (2\pi)^{3/2}} Exp\left\{-\frac{1}{2}\left(\frac{(v_x - u_x)^2}{w_x^2} + \frac{(v_y - u_y)^2}{w_y^2} + \frac{(v_z - u_z)^2}{w_z^2}\right)\right\}$$

Here, f(**v**) is normalized to the density, n, and centered at the velocity, **u** = {$u_x$, $u_y$, $u_z$}. The three thermal velocities are $w_j = \sqrt{(k_B T_j/m_s)}$, where j = (x, y, z), s = e or i, and $T_j$ is the temperature associated with direction i.  We define a thermal velocity vector, **w** = {$w_x$, $w_y$, $w_z$}.

All of the standard moment integrals in Eqns. (5) and (6) can be performed analytically, with the following results:

$(8a) \quad \int d^3\mathbf{v} f(\mathbf{v}) = n = $ density

$(8b) \quad \int d^3\mathbf{v} f(\mathbf{v})\mathbf{v} = n\mathbf{u} = $ particle flux of electrons or ions

$(8c) \quad U_{bulk} = \dfrac{nmu^2}{2}, \quad U_{therm} = \dfrac{nmw^2}{2}$

$$(8d) \quad \mathbf{P}_{RAM} = \text{nm}\mathbf{u}\mathbf{u}, \quad \mathbf{P} = \text{nm}\begin{bmatrix} w_x^2 & 0 & 0 \\ 0 & w_y^2 & 0 \\ 0 & 0 & w_z^2 \end{bmatrix}, \quad \text{or,} \quad \text{P}_{ls} = \text{nm}\delta_{ls}w_l^2 \text{ (no sum over l)}$$

$(8e) \quad \mathbf{Q}_{bulk} = \dfrac{n\mathbf{u}mu^2}{2}, \quad \mathbf{Q}_{enthalpy-l} = \dfrac{mn}{2}\left(u_l w^2 + 2u_l w_l^2\right) \text{ (no sum over l)}$

$(8f\ ) \quad \mathbf{Q}_{heatflux} = 0$

Note that the pressure tensor is diagonal in this coordinate system.  Such a coordinate system may be found for a non-diagonal pressure tensor in terms of its eigenvalues and eigenvectors.

The enthalpy flux vector has been calculated in terms of the pressure, as in Eqn. (5e), $\mathbf{Q}_{enthalpy} = \mathbf{u}\,\text{Tr}(\mathbf{P})/2 + \mathbf{u}\cdot\mathbf{P}$

The standard heat flux, $\mathbf{Q}_{heatflux}$, is zero for a tri-Maxwellian beam of the form in Eqn. (7) due to reflection symmetries which makes the heat flux integral vanish.

## 4. Multibeam moments

### 4.1 MULTIBEAM MOMENTS FOR f(**v**) = f₁ (**v**) + f₂ (**v**) + ... + f_N (**v**)

Consider a 3-D velocity distribution, f(**v**), consisting of N pieces, which will be referred to as *beams*:

(9) $f(v) = f_1 (v) + f_2 (v) + ... + f_N (v)$.

In the *standard* method for taking moments, a single mean flow velocity, **u**, is used, as in Eqns. (5) and (6) for the entire assembly, f(**v**) of beams. Such an approach is used in the typical calculation of on-board moments on a satellite, such as the Fast Plasma Instruments on board the MMS satellites as well as instruments on Cluster, THEMIS, etc.. It is usually applied without regard to whether or not the measured f(**v**) consists of one or more than one effectively disjoint "beams".

In the *multibeam* decomposition the *moments of each* beam are taken first (using flow velocities **u**_j of each *beam*, j). One then sums the moments over all beams. In general, the multibeam and standard decomposition methods give *different* results for the *decomposed* moments.

By contrast, the *undecomposed* moments, U, **Q**, and **T**, which are independent of flow velocities, are the *same* whether one uses f(v) or its equivalent, f₁(**v**) + f₂(**v**) + ... + f_N(**v**):

$$(10a) \quad \mathrm{U} \equiv \int d^3\mathbf{v} f(\mathbf{v})\left[\frac{mv^2}{2}\right] = \sum_j U_j, \quad U_j \equiv \int d^3v f_j(\mathbf{v})\left[\frac{mv^2}{2}\right]$$

$$(10b) \quad \mathbf{Q} \equiv \int d^3\mathbf{v} f(\mathbf{v})\left[\mathbf{v}\frac{mv^2}{2}\right] = \sum_j \mathbf{Q}_j, \quad \mathbf{Q}_j \equiv \int d^3\mathbf{v} f_j(\mathbf{v})\left[\mathbf{v}\frac{mv^2}{2}\right]$$

$$(10c) \quad \mathbf{T} \equiv \int d^3v f(v)\left[m\mathbf{vv}\right] = \sum_j \mathbf{T}_j, \quad \mathbf{T}_j \equiv \int d^3\mathbf{v} f_j(\mathbf{v})\left[m\mathbf{vv}\right]$$

The sum in Eqn. (10a) is a sum over the N beams of *undecomposed* beam energy densities, U_j; the sum in Eqn. (10b) is a sum over beams of undecomposed beam energy fluxes, **Q**_j; and the sum in Eqn. (10c) is a sum over beams of undecomposed individual beam stress tensors. It should be clear from Eqns. (10) that the *sum over undecomposed* moments *of each beam* gives the *same* result as the undecomposed moments of f(**v**). The undecomposed moments are therefore a robust and unambiguous characterization of the plasma, whether it consists of one or more beams.

The "*standard*" decomposition of U, **Q** and **T** based on the mean flow velocity **u**, is correctly given by Eqns. (5) and (6). The standard decomposed moments U_bulk, U_therm, **Q**_bulk, **Q**_heatflux, and **Q**_enthalpy, **T**_RAM and **T** are found, for example, in the moments file of the Fast Plasma Device (FPI) on each of the MMS spacecraft.

By contrast, in the alternative, *multibeam* decomposition one first decomposes the moments of each beam using that beam's mean velocity, $\mathbf{u}_j$ and then substitutes, $\mathbf{v} = \mathbf{u}_j + (\mathbf{v}_j - \mathbf{u}_j)$ in $mv^2/2$ and $mvv^2/2$ in the velocity integrals for $U_j$ and $\mathbf{Q}_j$ and $\mathbf{T}_j$ in Eqn. (10).

A key point in this paper is that each *multibeam decomposed* moment of the *assembly* of beams is given by the sum over beams, j, of the corresponding standard moment of each beam j,

$$(11a) \quad U_{bulk}^{MB} = \sum_{j=1}^{N} U_{j-bulk} \neq U_{bulk}, \quad U_{therm}^{MB} = \sum_{j=1}^{N} U_{j-therm} \neq U_{therm},$$

where, $U_{j-therm} = \dfrac{m}{2} \int d^3 \mathbf{v} f_j(\mathbf{v})(\mathbf{v} - \mathbf{u}_j)^2$

$$(11b) \quad \mathbf{Q}_{bulk}^{MB} = \sum_{j=1}^{N} \mathbf{Q}_{j-bulk} \neq \mathbf{Q}_{bulk}, \quad \mathbf{Q}_{heatflux}^{MB} = \sum_{j=1}^{N} \mathbf{Q}_{j-heatflux} \neq \mathbf{Q}_{heatflux}, \quad \mathbf{Q}_{enthalpy}^{MB} = \sum_{j=1}^{N} \mathbf{Q}_{j-enthalpy} \neq \mathbf{Q}_{enthalpy}$$

$$(11c) \quad \mathbf{P}^{MB} = \sum_{j=1}^{N} \mathbf{P}_j \neq \mathbf{P}, \quad \mathbf{P}_{RAM}^{MB} = \sum_{j=1}^{N} \mathbf{P}_{j-RAM} \neq \mathbf{P}_{RAM}$$

In general, none of the *multibeam* decomposed moments is equal to the corresponding *standard* decomposed moment.

Based on Eqns. (10) and (11) the *undecomposed* moments U, $\mathbf{Q}$ and $\mathbf{T}$ may therefore be decomposed into either a sum of standard moments or a sum of multibeam moments:

$$(12a) \quad U = U_{bulk}^{MB} + U_{therm}^{MB} = U_{bulk} + U_{therm}$$

$$(12b) \quad \mathbf{T} = \mathbf{P}_{RAM}^{MB} + \mathbf{P}^{MB} = \mathbf{P}_{RAM} + \mathbf{P}$$

$$(12c) \quad \mathbf{Q} = \mathbf{Q}_{bulk}^{MB} + \mathbf{Q}_{therm}^{MB} = \mathbf{Q}_{bulk} + \mathbf{Q}_{therm}, \text{ where,}$$

$$(12d) \quad \mathbf{Q}_{therm}^{MB} \equiv \mathbf{Q}_{heatflux}^{MB} + \mathbf{Q}_{enthalpy}^{MB} \text{ and } \mathbf{Q}_{therm} \equiv \mathbf{Q}_{heatflux} + \mathbf{Q}_{enthalpy}$$

The *thermal* energy density fluxes $\mathbf{Q}_{therm}$ and $(\mathbf{Q}^{MB})_{therm}$ have been defined in Eqn (12d) as sums of enthalpy flux and heat flux. Section (4.4) will show how Eqns. (12) determine the *false-thermal* parts of standard moments. First, we consider lower-order moments.

### 4.2 CONSERVATION OF PARTICLES AND OF PARTICLE FLUX

The zero-order-moment of $f(\mathbf{v}) = f_1(\mathbf{v}) + f_2(\mathbf{v}) + ... + f_N(\mathbf{v})$, yields conservation of particles,

$$13a) \quad \int \mathbf{d}^3 \mathbf{v} f(\mathbf{v}) / n = 1 = \eta_1 + \eta_2 + ... + \eta_N, \quad \eta_j \equiv n_j / n$$

Here, $n_j$ is the density of beam j, n is the mean density of the ensemble of beams and $\eta_j = n_j/n$ is the fractional density of each beam. The first-order moment of $f(\mathbf{v})$ yields conservation of particle flux. When divided by n, this becomes

(13b)   $\int \mathbf{d}^3\mathbf{v}\mathbf{v}f(\mathbf{v})/n = \mathbf{u} = \eta_1\mathbf{u}_1 + \eta_2\mathbf{u}_2 + ... + \eta_N\mathbf{u}_N,$

where $u_j$ is the velocity of the j-th beam and $\mathbf{u}$ is the mean velocity of the ensemble of beams. Each beam velocity, $\mathbf{u}_j$, in the weighted mean, $\mathbf{u}$, is multiplied by the fractional beam density, $\eta_j$.

Eqns. (13) illustrate that the beam velocities, $\mathbf{u}_j$, and the beam densities, $n_j$, are not completely independent of each other. Their relationship depends on the standard density, n and flow velocity, $\mathbf{u}$. We next consider a sum of tri-Maxwellian beams.

## 4.3 MULTIBEAM MOMENTS OF A SUM OF TRI-MAXWELLIANS

Consider an *analytic* velocity distribution consisting of a sum of N electron or N ion beams, $f(\mathbf{v}) = f_1(\mathbf{v}) + f_2(\mathbf{v}) + ... f_N(\mathbf{v})$, in which each beam, $f_j(\mathbf{v})$, is tri-Maxwellian:

(14)   $f_j(\mathbf{v}) = \dfrac{n_j}{w_{jx}w_{jy}w_{jz}(2\pi)^{3/2}} Exp\left\{-\dfrac{1}{2}\left(\dfrac{(v_x-u_{jx})^2}{w_{jx}^2} + \dfrac{(v_y-u_{jy})^2}{w_{jy}^2} + \dfrac{(v_z-u_{jz})^2}{w_{jz}^2}\right)\right\}$,   j = 1, 2, ..., N

Here $n_j$ is the density of beam j, $\mathbf{u}_j$ is its mean velocity, and $mw_{jx}^2 = T_{jx}$, $mw_{jy}^2 = T_{jy}$, $mw_{jz}^2 = T_{jz}$ are the three temperatures of beam j. We have taken all of the tri-Maxwellians to be co-aligned, with temperatures along the x, y and z axes. Consequently, all beam pressure tensor moments will here be diagonal. This of course, is not the most general possible sum of tri-Maxwellians but it will suffice to illustrate the concepts considered in this paper. More commonly treated is the even less general limit of a sum of *spherically symmetric* Maxwellians, each of temperature $T_j$, arrived at by setting $w_{jx} = w_{jy} = w_{jz} = \sqrt{(T_j/m)}$.

As in Eqns. (11), the *decomposed multibeam* moments of *any* multibeam $f(\mathbf{v})$ are given simply by the sum over beams, j, of the corresponding standard moments of each beam j. Since each beam is a tri-Maxwellian and the standard moments of a tri-Maxwellian are given by Eqns. (8), the *decomposed multibeam* moments of the sum of tri-Maxwellians can be expressed as,

(15a)   $U_{bulk}^{MB} = \dfrac{mn}{2}\sum_{j=1}^{N}\eta_j|\mathbf{u}_j|^2$,   $U_{therm}^{MB} = \dfrac{mn}{2}\sum_{j=1}^{N}\eta_j|\mathbf{w}_j|^2$

(15b)   $P_{ls}^{MB} = mn\delta_{ls}\sum_{j=1}^{N}\eta_j w_{jl}^2$,   (no sum over repeated l = x, y, z)

(15c)   $\mathbf{Q}_{bulk}^{MB} = \dfrac{mn}{2}\sum_{j=1}^{N}\eta_j\mathbf{u}_j|\mathbf{u}_j|^2$

(15d)   $\mathbf{Q}_{enthalpy-l}^{MB} = \dfrac{mn}{2}\sum_{j=1}^{N}\eta_j u_{jl}\left[|\mathbf{w}_j|^2 + 2w_{jl}^2\right]$   (no sum over repeated l = x, y, z)

(15e)   $\mathbf{Q}_{heatflux}^{MB} = 0$

Here the integer j labels *beams* while the integers l and s label vector and tensor *components* in the three Cartesian coordinates, {x, y, z}. The *multibeam* heat flux is zero because of velocity reflection symmetry in each tri-Maxwellian beam. We have defined a thermal *vector*, $\mathbf{w}_j \equiv \{w_{jx}, w_{jy}, w_{jz}\}$. In the *spherically symmetric limit*, $|\mathbf{w}_j|^2 = 3T_j/m$. In the *cold beam limit*, $w_j \rightarrow 0$, for all beams so the ensemble thermal energy, enthalpy flux and pressure go to zero.

## 4.4   HOW TO FIND *STANDARD* MOMENTS FROM *MULTIBEAM* MOMENTS

The decomposed moments obtained using the *standard* (e.g., FPI) mean-flow-velocity, **u**, and associated density, n, are given by the f(**v**) integrals in Eqns. (5) and (6).

First substitute f(**v**) = f₁(**v**) + ... + f_N(**v**) into Eqns. (5) and (6). Then, substitute [(**u_j**-**u**)+(**v**-**u_j**)] for δv ≡ (**v** - **u**) in each f_j(**v**) integral in each moment. Note that f_j(**v**) integrals proportional to (**v**-**u_j**) vanish because they are fluctuation moments. The standard energy density Eqns. (5a) then become,

$$(16a) \quad U_{bulk} = \frac{mn}{2}u^2, \quad U_{therm} = U_{therm}^{MB} + \frac{mn}{2}\sum_{j=1}^{N}\eta_j\left(\mathbf{u}-\mathbf{u}_j\right)^2 ,$$

the standard pressure and standard enthalpy flux Eqns. (5e) become,

$$(16b) \quad \mathbf{P} = \mathbf{P}^{MB} + mn\sum_{j=1}^{N}\eta_j\left(\mathbf{u}-\mathbf{u}_j\right)\left(\mathbf{u}-\mathbf{u}_j\right), \quad \mathbf{Q}_{enthalpy} = \frac{\mathbf{u}\,\mathrm{Tr}(\mathbf{P})}{2} + \mathbf{u}\cdot\mathbf{P} ,$$

and the standard bulk energy flux remains equal to Eqn. (5b):

$$(16c) \quad \mathbf{Q}_{bulk} = \frac{mn}{2}\mathbf{u}u^2$$

The standard heat flux in Eqn. (5c) requires a bit more algebra to relate to the multibeam heat flux:

$$(16d) \quad \mathbf{Q}_{heatflux} = \mathbf{Q}_{heatflux}^{MB} - \frac{nm}{2}\left\{\sum_{j=1}^{N}\eta_j\left[\mathbf{u}-\mathbf{u}_j\right]\left|\mathbf{u}-\mathbf{u}_j\right|^2 - 2\sum_{j=1}^{N}\eta_j\left[\mathbf{u}-\mathbf{u}_j\right]\left[\cdot\mathbf{P}_j + \frac{Tr\mathbf{P}_j}{2}\right]\right\}$$

### 4.4.1 STANDARD MOMENTS OF AN ASSEMBLY OF TRI-MAXWELLIANS

For the assembly of N co-aligned tri-Maxwellians, (U_therm)^MB and **P**^MB are known, so the *standard thermal* moments for the assembly are:

$$(17a) \quad U_{therm} = U_{therm}^{MB} + \frac{mn}{2}\sum_{j=1}^{N}\eta_j\left|\delta u_j\right|^2 = \frac{mn}{2}\left[\sum_{j=1}^{N}\eta_j\left(\left|\mathbf{w}_j\right|^2 + \left|\delta u_j\right|^2\right)\right], \quad \delta\mathbf{u}_j \equiv \mathbf{u}_j - \mathbf{u}$$

$$(17b) \quad \mathbf{P} = \mathbf{P}^{MB} + mn\sum_{j=1}^{N}\eta_j\delta\mathbf{u}_j\delta\mathbf{u}_j = \sum_{j=1}^{N}\mathbf{P}_j, \quad \mathbf{P}_j \equiv mn\eta_j\begin{bmatrix} w_{jx}^2+\left|\delta u_{jx}\right|^2 & \delta u_{jx}\delta u_{jy} & \delta u_{jx}\delta u_{jz} \\ \delta u_{jx}\delta u_{jy} & w_{jy}^2+\left|\delta u_{jy}\right|^2 & \delta u_{jz}\delta u_{jy} \\ \delta u_{jx}\delta u_{jz} & \delta u_{jz}\delta u_{jy} & w_{jz}^2+\left|\delta u_{jz}\right|^2 \end{bmatrix}$$

$$(17c) \quad \mathbf{Q}_{enthalpy} = \frac{\mathbf{u}\,\mathrm{Tr}(\mathbf{P})}{2} + \mathbf{u}\cdot\mathbf{P} = \mathbf{u}U_{therm} + \mathbf{u}\cdot\mathbf{P}$$

$$(17d) \quad \mathbf{Q}_{heatflux} = 0 + \frac{nm}{2}\left\{\sum_{j=1}^{N}\eta_j\delta\mathbf{u}_j\left|\delta u_j\right|^2\right\} + 2\sum_{j=1}^{N}\delta\mathbf{u}_j\left[\cdot\mathbf{P}_j + \frac{Tr\mathbf{P}_j}{2}\right]$$

There are both *true* and *false* thermal parts contained in the *standard* thermal energy density, heat flux and enthalpy flux of the assembly of N co-alignd tri-Maxwellian beams Although the *multibeam* heat flux, $(\mathbf{Q}_{\text{heatflux}})^{\text{MB}}$ of the assembly is zero, owing to the symmetry of each tri-Maxwellian beam, the *standard* heat flux of the assembly, $\mathbf{Q}_{\text{heatflux}}$, does *not* vanish, but is composed <u>entirely</u> of *false* thermal heat flux. Note that the dependence of $\mathbf{Q}_{\text{heatflux}}$ on $\mathbf{P}_j$ means that the *false* thermal heatflux contains thermal energy $|w_{jl}|^2$, where the index l is the Cartesian coordinate, l = x, y or z. However, in the cold beam limit $w_{jl} \rightarrow 0$, $\mathbf{Q}_{\text{heatflux}}$ does *not* go to zero.

### 4.4.2  MOMENTS OF EQUAL AND OPPOSITE MAXWELLIAN DISTRIBUTIONS

The above case is a generalization of the "false-thermal" standard energy density moment of two equal and opposite *cold* beams described in the introduction. For the two equal and opposite cold beams in Fig. (1), the mean flow velocity, u = 0, and, from Eqns. (15) and (16),

$$n_1 = n_2 = \frac{n}{2}, \quad \mathbf{u}_1 = \mathbf{u}_0, \quad \mathbf{u}_2 = -\mathbf{u}_0, \quad \mathbf{u} = 0$$

$$U_{bulk} = 0, \quad U_{therm} = \frac{m}{2}\left(n_1 u_0^2 + n_1 u_0^2\right) = mn_1 u_0^2$$

$$U_{bulk}^{MB} = \frac{m}{2}\left(n_1 u_0^2 + n_1 u_0^2\right) = mn_1 u_0^2, \quad U_{therm}^{MB} = 0$$

Thus, the standard (FPI) moments identify the energy of an assembly of cold beams as totally "thermal," whereas multibeam moments correctly identify the system energy as composed of bulk kinetic energy only. This is what is termed here as a *false* thermal component.

In general, $U_{\text{thermt}}$ will be a mixture of real and false thermal energy densities. If the two cold beams in Fig. 1 and Table 1 are replaced by two Maxwellian beams, each with isotropic temperature, $T = mw^2$, Eqns. (17) still hold, except for the "thermal" energy densities expressions which are changed to:

$$U_{therm} = \frac{mn\left\{w^2 + u_0^2\right\}}{2} \qquad U_{therm}^{2-beam} = \frac{mnw^2}{2} = U_{therm} - \left(\frac{mnu_0^2}{2}\right)$$

Table (2), below, compares the standard energy density moments to the multibeam energy density moments for this example of two beams.

**Table 2:  STANDARD VS. MULTIBEAM energy density moments for two equal (density) and opposite warm beams (e or i)**

| *Standard* energy moments | | *Multibeam* energy moments | |
|---|---|---|---|
| $U_{bulk}$ | 0 | $(U_{bulk})^{2\text{-beam}}$ | $\dfrac{mnu_0^2}{2}$ |
| $U_{therm}$ | $\dfrac{mn\left\{w^2 + u_0^2\right\}}{2}$ | $(U_{therm})^{2\text{-beam}}$ | $\dfrac{mnw^2}{2}$ |
| $U = U_{bulk} + U_{therm}$ | $\dfrac{mn\left\{w^2 + u_0^2\right\}}{2}$ | $U = (U_{bulk})^{2\text{-beam}} + (U_{therm})^{2\text{-beam}}$ | $\dfrac{mn\left\{w^2 + u_0^2\right\}}{2}$ |

### 4.5 *FALSE-THERMAL* PARTS OF DIMENSIONLESS STANDARD MOMENTS

Thus, *standard* thermal energy moments may contain *false-thermal* parts, whereas the corresponding *multibeam* thermal energy moments are entirely *true-thermal*.

In general, false-thermal *parts* of standard moments are obtained by subtracting from each standard moment the corresponding multibeam moment. Hence, the false-thermal moments, $\Delta U$, $\Delta \mathbf{P}$, and $\Delta \mathbf{Q}$ are defined as:

(18a) $\quad \Delta U \equiv U_{therm} - U_{therm}^{MB}$

(18b) $\quad \Delta \mathbf{P} \equiv \mathbf{P} - \mathbf{P}^{MB}$

(18c) $\quad \Delta \mathbf{Q} \equiv \mathbf{Q}_{therm} - \mathbf{Q}_{therm}^{MB}$

Since Eqn.(12a), for example, may be rewritten as $U_{therm} - U_{therm}^{MB} = U_{bulk}^{MB} - U_{bulk}$ , Eqns. (12) allow us to replace the difference moments in Eqn. (18) by differences of the much simpler *bulk* moments:

(19a) $\quad \Delta U \equiv U_{bulk}^{MB} - U_{bulk}$

(19b) $\quad \Delta \mathbf{P} \equiv \mathbf{P}_{RAM}^{MB} - \mathbf{P}_{RAM}$

(19c) $\quad \Delta \mathbf{Q} \equiv \mathbf{Q}_{bulk}^{MB} - \mathbf{Q}_{bulk}$

First consider energy density moments. Define *dimensionless energy units* by dividing *all* energy densities by a *normalization* standard bulk energy density, $U_{norm} = mnu_n^2/2$. We allow for the possibility that the normalization flow velocity moment, $\mathbf{u}_n$ may (or may not) differ from the standard flow velocity moment, $\mathbf{u}$, of the given $f(\mathbf{v})$.

The *standard and multibeam bulk energy density moments* are independent of the exact shape of $f(\mathbf{v}) = f_1(\mathbf{v}) + ... + f_N(\mathbf{v})$ and of the shape of the individual beam distributions, $f_j(\mathbf{v})$. These moments depend only on densities and mean velocities of $f(\mathbf{v})$, $(n, \mathbf{u})$, and of $f_j(\mathbf{v})$, $(n_j, \mathbf{u}_j)$, so that the *false-thermal energy density moment* is

20a) $\quad \Delta U \equiv U_{bulk}^{MB} - U_{bulk} \equiv \dfrac{1}{u_n^2}\left[ \sum_{j=1}^{N} \eta_j u_j^2 - u^2 \right], \quad \eta_j = \dfrac{n_j}{n}$

Two equivalent alternate expressions for the *dimensionless* $\Delta U$ are

(20b) $\quad \Delta U = \dfrac{1}{u_n^2}\sum_{j=1}^{N}\eta_j\left( \mathbf{u}_j - \mathbf{u} \right)^2$

(20c) $\quad \Delta U = \dfrac{1}{u_n^2}\sum_{\substack{i,j=1 \\ i<j}}^{N}\eta_i\eta_j\left( \mathbf{u}_i - \mathbf{u}_j \right)^2$

Eqn. (20b) has N terms. It is equivalent to Eqn. (20a) as one can see by explicitly squaring $(\mathbf{u}-\mathbf{u}_j)^2$ in each term and summing. In this form, $\Delta U$ is manifestly frame-independent, since the beam velocities, $\mathbf{u}_j$ are all measured from the mean flow velocity, $\mathbf{u}$. The right side of Eqn. (20c) depends on *differences* of beam velocities, $(\mathbf{u}_i-\mathbf{u}_j)$ so it too is explicitly frame-independent. In this format there are N(N-1)/2 terms.

Any of the Eqns. (20) may be used to *calculate* the false-thermal energy density, $\Delta U$, in terms of zero-order and first-order moments of $f(\mathbf{v})$ and of zero-order and first-order moments of each of the beams, $f_j(\mathbf{v})$. However, there is another possible use of $\Delta U$ which will be demonstrated in a later paper dealing with methods for explicit multibeam moments of *measured* $f(\mathbf{v})$. From Eqn. (19a), the multibeam thermal energy density $(U_{therm})^{MB}$ is,

(21)    $U_{therm}^{MB} = U_{therm} - \Delta U$

Thus, for a given (i.e., measured) $f(v)$, and a known (i.e., measured) $U_{therm}$ the decomposed multibean energy density, $(U_{therm})^{MB}$ can be found by sutracting $\Delta U$. This is usually easier than calculating $(U_{therm})^{MB}$ by summing all of the beam thermal energy densities (as in Eqn. (11a))

Consistent with the treatment leading to the multibeam energy density, we next treat the vector, *false energy flux, $\Delta \mathbf{Q}$*. Here, all energy *fluxes* are rendered dimensionless by dividing by $mnu_n^3/2$:

(22)    $\mathbf{Q}_{bulk}^{MB} = \mathbf{Q}_{bulk} + \Delta \mathbf{Q}, \quad \Delta \mathbf{Q} \equiv \dfrac{1}{u_n^3}\left[\displaystyle\sum_{j=1}^{N} \eta_j \mathbf{u}_j u_j^2 - \mathbf{u}u^2\right]$

By the same reasoning that led to the "*thermal*" multibeam *energy* moment in Eqn. (21), we can express the dimensionless "*thermal*" multibeam *energy flux* vector moment as,

(23)    $\mathbf{Q}_{therm}^{MB} = \mathbf{Q}_{therm} - \Delta \mathbf{Q}$

Hence, if we know the standard moment $\mathbf{Q}_{therm}$ for $f(\mathbf{v})$, we can find $(\mathbf{Q}_{therm})^{MB}$ by subtracting the "false thermal" part, $\Delta \mathbf{Q}$. Adding Eqns. (22) and (23) verifies that either way the moments are taken (standard method or multibeam method) one obtains the same undecomposed $\mathbf{Q}$. The same is true of the undecomposed energy density, U, as is evident when $\Delta U$ is eliminated from Eqns. (18a) and (21).

One can use a similar strategy to find the multibeam pressure tensor, $\mathbf{P}^{MB}$, from the known standard pressure tensor, $\mathbf{P}$ by introducing the RAM pressure tensor moment, $\mathbf{P}_{RAM}$ and stress tensor moment, $\mathbf{T}$, as in Eqn. (12). Here, $\mathbf{T} = \mathbf{P} + \mathbf{P}_{RAM}$, has the same value for standard or multibeam pressure moments. Dimensionless pressures are obtained by dividing all pressures by $nmu_n^2$. If $\Delta \mathbf{P}$ is defined as $\Delta \mathbf{P} \equiv \mathbf{P}_{RAM} - \mathbf{P}_{RAM}^{MB}$, then, by the same arguments as above,

24)    $\mathbf{P}^{MB} = \mathbf{P} - \Delta \mathbf{P}$

25)    $\Delta \mathbf{P} \equiv \dfrac{1}{u_n^2}\left[\displaystyle\sum_{j=1}^{N} \eta_j \mathbf{u}_j \mathbf{u}_j - \mathbf{u}\mathbf{u}\right]$

Note that the multibeam pressure moment tensor, $\mathbf{P}^{MB}$, *contains* the scalar multibeam energy density, $U_{therm}^{MB}$, since, in *dimensionless units*, the *trace* of $\mathbf{P}^{MB}$, $\text{Tr}[\mathbf{P}^{MB}] = U_{therm}^{MB}$.

When $\mathbf{u}$, $\mathbf{P}$, and $\mathbf{Q}$ are *known* standard moments of a known or measured $f(\mathbf{v})$, one need only find the density and velocity $(n_j, u_j)$ of each port Equation, $f_j(\mathbf{v})$ in order to evaluate $\Delta P$ and $\Delta \mathbf{Q}$ and, from them, $\mathbf{P}^{MB}$, $(U_{therm})^{MB}$ and $(\mathbf{Q}_{itherm})^{MB}$. (See Table 3). This is a more effective strategy than calculating a sum of higher order beam thermal moments, as in the definitions of $\mathbf{P}^{MB}$, $(U_{therm})^{MB}$ and $(\mathbf{Q}_{itherm})^{MB}$ given in Eqns. (11).

**Table 3:** **Formulas** for determining multibeam thermal moments from standard thermal moments. (dimensionless unit scaling factor shown in third column).

| f(v) moment | False-thermal part of standard moment | Units | Multibeam moment |
|---|---|---|---|
| **Energy Density** | $\Delta U \equiv \dfrac{1}{u_n^2}\left[\sum_{j=1}^{N}\eta_j u_j^2 - u^2\right]$ | $\dfrac{mnu_n^2}{2}$ | $\boxed{U_{therm}^{MB} = U_{therm} - \Delta U}$ |
| **Pressure tensor** | $\Delta \mathbf{P} \equiv \dfrac{1}{u_n^2}\left[\sum_{j=1}^{N}\eta_j \mathbf{u}_j \mathbf{u}_j - \mathbf{uu}\right]$ | $mnu_n^2$ | $\boxed{\mathbf{P}^{MB} = \mathbf{P} - \Delta \mathbf{P}}$ <br> $Tr\mathbf{P} = U_{therm}, \quad Tr\mathbf{P}^{MB} = U_{therm}^{MB}$ |
| **Energy Flux vector** | $\Delta \mathbf{Q} \equiv \dfrac{1}{u_n^2}\left[\sum_{j=1}^{N}\eta_j \mathbf{u}_j u_j^2 - \mathbf{uu}^2\right]$ | $\dfrac{mnu_n^3}{2}$ | $\boxed{\mathbf{Q}_{therm}^{MB} = \mathbf{Q}_{therm} - \Delta \mathbf{Q}} \quad \mathbf{Q}_{therm} \equiv \mathbf{Q}_{enthalpy} + \mathbf{Q}_{htflux},$ <br> $\mathbf{Q}_{enthalpy} \equiv \hat{\mathbf{u}}Tr\mathbf{P} + 2\hat{\mathbf{u}} \cdot \mathbf{P}$ |

## 4.6 MULTIBEAM ENERGY TRANSPORT EQUATIONS

The multibeam energy moments satisfy their own multibeam space-time energy transport equations, analogous to the normal energy transport equations, Eqns. (1):

26a) $\partial_t U_{bulk}^{MB} + \nabla \cdot \mathbf{Q}_{bulk}^{MB} = \mathbf{J} \cdot \mathbf{E} - \sum_{i=1}^{N}\mathbf{u}_i \cdot \nabla \cdot \mathbf{P}_i^{MB}$

26b) $\partial_t U_{thermal}^{MB} + \nabla \cdot \mathbf{Q}_{thermal}^{MB} = \sum_{i=1}^{N}\mathbf{u}_i \cdot \nabla \cdot \mathbf{P}_i^{MB}, \quad \mathbf{Q}_{thermal}^{MB} \equiv \mathbf{Q}_{heatflux}^{MB} + \mathbf{Q}_{enthalpy}^{MB}$

Here $\mathbf{J}$ is the system current equal to the sum over all N beam currents. The transfer of energy between multibeam bulk energy and multibeam thermal energy is governed by the sum over beams of the work done by each beam's pressure against that beam's flow velocity. Note that the multibeam pressure which appears in the multibeam *force-momentum* eqn. does not appear in the multibeam energy transport equations because the work done by the pressure in the multibeam energy transport equation must be computed beam by beam and then summed.

# 5.  Summary and conclusions

## 5.1 MULTIBEAM VELOCITY DISTRIBUTIONS AND FALSE THERMAL MOMENTS

In this paper we have addressed issues that arise when taking energy moments of effectively disjoint particle velocity distributions, f(**v**), which can be expressed as a sum of N beams, f(**v**) = f₁(**v**) + … + f_N(**v**).

The *standard* moments of any f(**v**) are based on the assumption that there is only *one* flow velocity, **u**, associated with f(**v**).  Examples of standard *coherent* moments are the bulk kinetic energy density, $U_{bulk} = nmu^2/2$, of the system and the bulk kinetic energy density flux vector, **Q**_bulk = n**u**mu²/2 of the system.  Moments which depend on correlations of fluctuations (**v** − **u**) are said to be *incoherent or "thermal"*.  Examples are the standard thermal energy density, $U_{thermal} = \int d^3 \mathbf{v} f(\mathbf{v}) \, m|\mathbf{v} - \mathbf{u}|^2/2$, and the standard thermal energy flux vector, **Q**_thermal, defined, in our convention, to be the sum of the enthalpy and heat flux vectors.  There are separate space-time transport equations for the coherent and the incoherent energy moments (Eqns. (1)).  In this paper we do not address the space-time evolution of energy moments of electrons or ions but rather consider their properties averaged over short time periods and small spatial regions.

*Standard* moments of multibeam velocity distributions f(**v**) = f₁(**v**) + … + f_N(**v**) tend to inflate U_thermal and change the direction and magnitude of **Q**_thermal by amounts we call the *false parts* of these standard thermal moments, ΔU_thermal and Δ**Q**_thermal.  A simple example is the false thermal energy, ΔU_thermal, of a pair of cold beams.  In this case ΔU_thermal arises from the standard-moment misidentification of their center-of-mass-frame kinetic energy as thermal energy.

There is an alternate way of taking moments well-suited for multibeam velocity distributions of form f(**v**) = f₁(**v**) + … + f_N(**v**), which we call *multibeam* moments.  *Multibeam* thermal energy density moments, $U_{thermal}^{MB}$ and energy density flux moments, $\mathbf{Q}_{thermal}^{MB}$, do not have false parts. They are calculated by taking a *standard* thermal moment of *each beam* and summing over beams.  For example, the multibeam thermal energy density moment is given by,

$$U_{thermal}^{MB} = \sum_{j=1}^{N} \int f_j(\mathbf{v}) m \, | \, \mathbf{v} - \mathbf{u}_j \, |^2 \, /2$$

The false part of a thermal moment is the difference between the standard and multibeam thermal moments. For example,  $\Delta U_{thermal} = U_{thermal} - U_{thermal}^{MB}$,  and,  $\Delta \mathbf{Q}_{thermal} = \mathbf{Q}_{thermal} - \mathbf{Q}_{thermal}^{MB}$. Coherent multibeam moments are calculated in the same way.  For example, the multibeam kinetic energy flux is given by,

$$U_{bulk}^{MB} = \sum_{j=1}^{N} mn_j \, | \, \mathbf{u}_j \, |^2 \, /2$$

The time-evolution of the *multibeam* coherent and thermal energy densities is governed by a pair of space-time energy equations (Eqns. (26)) similar to those for the normal bulk and thermal energy densities. Energy transfer between coherent and thermal multibeam energy densities is now governed by the sum over beams of the work done by the multibeam pressure density force of each beam on or against the beam's flow velocity, **u**_j.

To illustrate the advantages of taking multibeam energy moments the method was applied to a system of N co-axial tri-Maxwellians. Due to symmetry, none of the N tri-Maxwellians had a standard heat flux moment. However, the standard heat flux moment of the sum of beams, $f(\mathbf{v}) = f_1 + \ldots + f_N$ was non-zero. Hence *all* of the standard heat flux of $f(\mathbf{v})$ is *false and* the multibeam heat flux moment of $f(\mathbf{v})$ is zero.

For the special case of a pair of equal and opposite cold beams the normal thermal energy density is nonzero, while the multibeam thermal energy density is zero.

### 5.2 UNDECOMPOSED ENERGY MOMENTS

There is still another way of taking energy moments of $f(\mathbf{v})$, this time with no reference to either a single beam flow velocity or to multiple beam flow velocities and hence no distinction between coherence or incoherence of energy moments.

An example is the overall energy moment, $U = \int d^3\mathbf{v} f(\mathbf{v}) m v^2/2$. If $v^2$ is replaced by $[(\mathbf{v}-\mathbf{u}) + \mathbf{u}]^2$, U decomposes into a sum of the *normal* coherent (bulk kinetic) energy density and incoherent (thermal) energy density, $U = U_{bulk} + U_{thermal}$. For that reason, we have referred to U as the *undecomposed* energy density. For a system of N beams the substitution of $f = f_1 + \ldots + f_N$ in the integral for U, followed by a replacement of $v^2$ by $[(\mathbf{v}-\mathbf{u_j}) + \mathbf{u_j}]^2$ in each beam integral, j, leads to the decomposition of U into a sum of coherent and incoherent multibeam energy densities,
$U = U_{bulk}^{MB} + U_{thermal}^{MB}$.

Therefore, regardless of whether $f(\mathbf{v})$ consists of one or more beams, the undecomposed energy moments are meaningful and invariant under which method of decomposition is employed. This is true for the undecomposed energy flux vector, $\mathbf{Q}$, as well as for U. The space-time energy transport equation for U and $\mathbf{Q}$ is given by Eqn. (2a).

Analysis of *measured* velocity distributions, $f(\mathbf{v})$, using moments such as U and $\mathbf{Q}$ has the advantage that undecomposed energy moments are well-defined and meaningful for single-peaked, multi-peaked, disjoint or arbitrarily-shaped $f(\mathbf{v})$, far from thermal equilibrium. The disadvantage is that the energy and energy flux cannot be interpreted as "coherent" or "incoherent" (i.e., *bulk* or *thermal*).

One interesting and important consequence of the invariance of U and $\mathbf{Q}$ is that false thermal parts of normal thermal moments which are absent in multibeam *thermal* moments appear in the corresponding multibeam *bulk* moments. As one example, false normal thermal energy density in an assembly of beams will appear as an increase in (multibeam) bulk kinetic energy density. Equivalently, false temperature should be reinterpreted as increased kinetic energy per particle in the assembly of beams. In another example, false normal heat flux in a distribution consisting of a sum of beams (e.g., N tri-Maxwellians) will show up as extra enthalpy flux and kinetic energy flux in the collection of beams. Since the fluxes are vectors the more precise statement is
$\mathbf{Q}_{heatflux} - \mathbf{Q}_{heatflux}^{MB} = \left( \mathbf{Q}_{enthalpy}^{MB} + \mathbf{Q}_{bulk}^{MB} \right) - \left( \mathbf{Q}_{enthalpy} + \mathbf{Q}_{bulk} \right)$.

We have shown that the invariance of undecomposed energy moments leads to a simplification in the determination of the false thermal parts of the normal thermal energy and normal energy flux moments. For example, we may use the invariance of U and **Q** to rewrite the false normal thermal energy density and the false normal thermal energy flux in terms of

$\Delta U_{thermal} = U_{bulk}^{MB} - U_{bulk}$ and $\Delta \mathbf{Q}_{thermal} = \mathbf{Q}_{bulk}^{MB} - \mathbf{Q}_{bulk}$. Hence, the false parts of these normal thermal moments can be expressed entirely in terms of coherent beam velocities and beam densities. This will form the basis of the so-called "visual method" for finding false thermal moments of measured f(**v**) to be be described in a future paper.

### 5.3 PRESSURE MOMENTS

In addition to U and **Q**, another undecomposed moment we have treated is the particle stress tensor, $\boldsymbol{T} = \int d^3\mathbf{v} f(\mathbf{v}) m \ \mathbf{v}\mathbf{v}$, whose divergence appears in the particle momentum equation (Eqn. (6b)). $\boldsymbol{T}$ decomposes into the normal "thermal" pressure tensor moment, $\boldsymbol{P}$ and the RAM pressure tensor, $\mathbf{P}_{RAM} = m\mathbf{u}\mathbf{u}$ when both **v**'s in the dyadic **vv** are replaced by $\mathbf{v} = (\mathbf{v} - \mathbf{u}) + \mathbf{u}$. It also decomposes into the multibeam pressure tensor, $\boldsymbol{P}^{MB}$, and the RAM pressure tensor,

$\boldsymbol{P}_{RAM}^{MB} = \sum_{=1,j}^{N} mn_j \left( \mathbf{u}_j \mathbf{u}_j \right)$, when f(**v**) is replaced by (f$_1$(**v**) + … + f$_N$(**v**)) and **v** is expressed as

$\mathbf{v} = (\mathbf{v} - \mathbf{u}_j) + \mathbf{u}_j$ in the RAM pressure of each of the N beams. Once again, there wil be a *false* (tensor) part of the normal thermal pressure tensor moment, $\boldsymbol{P}$, in a multibeam f(**v**), given by the difference, $\Delta \boldsymbol{P} = \boldsymbol{P} - \boldsymbol{P}^{MB}$. False pressure will show up in the multibeam RAM pressure tensor due to the invariance of $\boldsymbol{T}$.

### 5.4 SIGNIFICANCE

Recent high-resolution simulations and measurements of electron and ion velocity distributions, f(**v**), during magnetic reconnection have revealed that f(**v**) can be very far from thermal equilibrium, quite complex, and often effectively disjoint. Hence the interpretation of particle energetics sometimes requires new tools although we are still wedded to using long-standing familiar ones such as fluid concepts and fluid models. Kinetic theory modeling of processes in collisionless plasmas is increasingly used to understand space physics measurements, especially through particle-in-cell simulations. Vlasov equations contain particle and field physics which is not present in fluid equations. Velocity moments of the Vlasov equation are often the basis for deriving fluid variables and equations.

In this paper we retain this framework but introduce a *multibeam* method of taking velocity moments which is sometimes more appropriate than the "*normal*" method of taking velocity moments for interpreting distributions, f(**v**), which are far from equilibrium and effectively disjoint. In the *normal* method of taking moments it is assumed that there is only one overall flow velocity associated with f(**v**) and the energy moments of f(**v**) are either coherent (bulk kinetic energy, bulk energy flux, RAM pressure, etc.) or incoherent (pressure, thermal energy density, enthalpy flux, heat flux, etc.). This can lead to counterintuitive results, such as the *false* normal thermal energy in a system, f(**v**) of two cold equal and opposite cold electron beams. The remedy is well known – consider each electron beam to be a separate *species*. When the system thermal energy is calculated for each species separately and then summed, the system energy is found to be entirely kinetic rather than entirely thermal.

The underlying approximation of the multibeam method is to express f($\mathbf{v}$) as a sum of distributions, which we call beams, so that for electrons or for ions, f($\mathbf{v}$) = $f_1$($\mathbf{v}$) + … + $f_N$($\mathbf{v}$). We can think of each beam as a separate subspecies within the N electrons or the N ions. Each beam has a centroid velocity, $\mathbf{u}_j$, a density, $n_j$, etc. Multibeam energy moments of f($\mathbf{v}$) are constructed by taking normal energy moments of each beam and then summing over beams. Thus, multibeam moments can again be either coherent (multibeam bulk kinetic energy, multibeam bulk energy flux, multibeam RAM pressure, etc.) or incoherent (multibeam pressure, multibeam thermal energy density, multibeam enthalpy flux, multibeam heat flux, etc.). *Multibeam* incoherent moments do not contain false thermal parts. However, any of the *normal* incoherent energy moments can contain false thermal parts which actually should be reassigned (whether scalar vector or tensor) to a corresponding multibeam coherent moment.

This has practical significance for *normal* fluid interpretations of *multibeam* f($\mathbf{v}$) measured by spacecraft during magnetospheric reconnection or found in PIC simulations. Some hypothetical scenarios illustrating the differences in interpretation when taking multibeam moments of a multibeam electron distribution, f($\mathbf{v}$), rather than taking normal moments are:

- o  If an elevated normal electron *temperature* T = ($U_{thermal}/n$) measured by an MMS spacecraft over a 30 ms time interval is interpreted as due to a *heating process* and the electron velocity distribution, f($\mathbf{v}$) measured by the FPI instrument is found to consist of disjoint electron beams, the multibeam electron temperature will be lower, and the multibeam bulk kinetic energy per particle higher by the same amount. Although the normal temperature suggests electron heating has occurred, the multibeam temperature reveals that some of the higher electron energy is coherent rather than incoherent, arising from electron acceleration to a higher bulk kinetic energy. The relative magnitudes of thermal energy density vs bulk kinetic energy remain to be studied for specific measurements.

- o  Suppose a measured electron distribution, f($\mathbf{v}$), consists of a (high-energy) crescent-shaped beam and a broad background with its own small flow velocity. The magnitude and direction of the (measured) *normal* heat flux moment of f($\mathbf{v}$) may differ from the magnitude and direction of the *multibeam* heat flux. In other words, the normal electron heat flux vector may have a false part which multibeam energy flux moments reveal to actually be carried by the electron enthalpy flux vector and the bulk electron kinetic energy flux vector.

It is worth restating and discussing the assumptions and limitations of the approach developed here for taking energy moments of non-equilibrium particle distributions, f($\mathbf{v}$). A number of these issues will be addressed by work currently in progress.

The multibeam moment method described here applies to non-equilibrium velocity distributions, f(**v**), essentially at one time and place, which can be approximated as a sum of sub-distributions, $f_1(\mathbf{v}) + ... + f_N$, referred to as "beams." While an analytic example has been given of a sum of tri-Maxwellians, it has not been demonstrated in this paper when or if an arbitrary *measured* non-equilibrium f(**v**) can be expressed in this manner. One strategy would be to approximate f(**v**) as a sum of parametrized analytic sub-distributions (not necessarily tri-Maxwellians) and to determine the parameters by a least-squares fit to the measured f(**v**). There are preliminary indications that this method can be useful and can be adapted to treat high-energy distributed velocity-space "clouds" or "haloes" as particular sub-distributions (beams) in the sum.

Aside from the question of uniqueness of a beam decomposition for a given N it is not always clear how many beams, N, should be included in the sum. In the theoretical limit in which N equals the number of *particles* in f(**v**), one enters the microscopic kinetic regime, where there is not even a concept of fluid variables such as temperature. However, published work on machine learning suggests a single digit-number of beams, N, is reasonable (DuPuis, et al, 2020).

The least squares method and other strategies for expressing ion distributions measured during magnetic reconnection in the magnetosphere (e.g., Fig. (2)) have been presented in papers at conferences, and is about to be submitted for publication.

Although multibeam space-time energy transport equations have been developed (Eqns. (26)), the present paper does not consider particle energy transport in space and time. The multibeam fluid moments taken here are of an Eulerian multibeam distribution, f(**v**), at one space-time point (or an average over at most a small range of such points). The study of *processes* such as multibeam particle energization and space-time energy transfer are beyond the scope of the moment methods discussed here. Nevertheless, *undecomposed* energy moments discussed in this paper may be useful in following spacecraft-measured Eulerian f(**v**) in time along the spacecraft orbit, especially when multiple beams appear to merge and split and when the distinction between coherent and incoherent moments is not essential.

Considerable insight is gained into the *origins and energization* of beams from *Lagrangian* studies based on kinetic theory (i.e., simulations), such as the particle trajectories traced in Fig. (3). For example, a pair of beams may be two-stream unstable only so long as the space-time particle trajectories remain crossed at the same spatial point. Even if instability does occur, the spacecraft must remain at that crossing point long enough for it to be observed.

In conclusion, care must be taken when applying fluid concepts to highly non-Maxwellian particle velocity distributions, especially when they are disjoint.

**Acknowledgments**

We wish to thank Dr. Steven Schwartz for helpful conversations leading to this paper.
JPE is supported by UKRI (STFC) grant ST/N000692/1.
M. Goldman, D. Newman and G. Lapenta are supported by NASA Grant NNX08AO84G

.